\documentclass{article}

\usepackage[english]{babel}

\usepackage[letterpaper,top=2cm,bottom=3.6cm,left=3cm,right=3cm,marginparwidth=1.75cm]{geometry}

\usepackage{amsmath, amssymb}
\usepackage{graphicx}
\usepackage[colorlinks=true, allcolors=blue]{hyperref}
\usepackage{placeins}
\usepackage{subcaption}
\usepackage[inkscapelatex=false]{svg}
\usepackage{float}
\usepackage{algpseudocode}
\usepackage{algorithm}
\usepackage{tikz}
\usetikzlibrary{positioning, arrows.meta, calc}
\usepackage{overpic}
\usepackage{fancyhdr}
\usepackage[
backend=biber,
sorting=none]{biblatex}
\title{\textbf{Short Ticketing Detection Framework Analysis Report}}
\author{Independent analysis provided via Imperial Consultants by: \\ Yuyang Miao, Huijun Xing, Danilo P. Mandic and Tony G. Constantinides\\
AIDA Lab, Imperial College London}
\date{June 30, 2025}

\begin{document}

\maketitle
\thispagestyle{fancy} 

\vspace{2cm}
\begin{center}
\large
\textbf{Abstract}
\end{center}
\vspace{0.5cm}

This report presents a comprehensive analysis of an unsupervised multi-expert machine learning framework for detecting short ticketing fraud in railway systems. The study introduces an A/B/C/D station classification system that successfully identifies suspicious patterns across 30 high-risk stations. The framework employs four complementary algorithms: Isolation Forest, Local Outlier Factor, One-Class SVM, and Mahalanobis Distance. Key findings include the identification of five distinct short ticketing patterns and potential for short ticketing recovery in transportation systems.
\footnote{This report represents the independent opinion of the author(s) and may be subject to copyright.}

\vspace{2cm}
\begin{center}
\large
\textbf{Keywords}
\end{center}
\vspace{0.5cm}

Short ticketing, unsupervised learning, railway systems, anomaly detection, machine learning, multi-expert

\newpage

\begin{center}
\Large\textbf{Table of Contents}
\end{center}
\vspace{1cm}

\tableofcontents
\thispagestyle{fancy} 

\newpage

\section{Executive Summary}
Each year, fare evasion costs the UK railway system approximately 240 million pounds~\cite{facit2025reclaiming} with short ticketing, where passengers buy tickets for shorter, cheaper journeys but travel beyond the permitted destination, representing a specific and often undetected aspect of the broader issue. A simple but practical example would be: a passenger travelling from Seaside Station to International Terminus Station via Commuter Hub Station and Financial District Station might purchase two separate tickets (Seaside Station to Commuter Hub Station, and Financial District Station to International Terminus Station) instead of the complete journey ticket, potentially saving money while committing ticket fraud leading to revenue loss for the Train Operating Companies (TOCs).

To solve this problem, this comprehensive report provides an in-depth analysis of the short ticketing detection framework developed by researchers Yuyang Miao and Huijun Xing at Imperial College London. This study represents an unsupervised machine learning approach. This work is based on a dataset collected from the UK railway system, including 100 stations' entry and exit data for seven days, with approximately 6.5 million trials of records. \footnote{The station names have been anonymised within the rest of this report. The following aliases are used in this report: Seaside Station, International Terminus Station, Commuter Hub Station, Financial District Station, Airport Terminal Station, Downtown Station, and West Interchange Station.}

The research introduces an  A/B/C/D station classification system that categorizes each station as Actual Entry (A), Declared Destination (B), Declared Origin (C), or Actual Exit (D). This framework enables the identification of gaps between passengers' declared travel journeys and their actual travel journeys, which is the fundamental mechanism underlying short ticketing behaviour.

\subsection{Critical Findings}

The framework successfully identified 30 high-risk stations showing highly unusual activity. The Airport Terminal Station was flagged as the highest-risk location, exhibiting 'Ghost Station' behaviour. Here, a large majority of transactions were for entries or exits only, with no complete journeys being registered. This station's activity was a significant statistical departure from normal patterns. Similarly, a gate line at Downtown Station ranked as the second-highest risk. It demonstrated 'Black-Hole' behaviour, where most of the recorded taps were for exits, strongly suggesting a high level of fare evasion. Analysis confirmed that both stations are operating far outside of normal, expected patterns.

\subsection{Critical Methodological Innovation}
The research employs four complementary unsupervised learning algorithms, namely the Isolation Forest, Local Outlier Factor, One-Class SVM, and Mahalanobis Distance, combined through an adaptive weighting system that dynamically adjusts each method's contributions based on cross-correlation analysis. This multi-expert approach keeps a trade-off between accuracy and stableness for the analysis.

\subsection{Operational Impact}
The framework identifies five distinct short ticketing types: Ghost Station, Black-Hole, Fake-Origin, Function-Loss, and Micro-Trap behaviors. Each pattern represents different exploitation mechanisms, enabling a specific operation possibility. The quantitative risk assessment allows railway TOCs to prioritize limited resources effectively, potentially recovering millions in lost revenue.

\subsection{Strategic Implications}
This research establishes a new method for data-driven, unsupervised detection of short ticketing. The methodology can identify complex fraud patterns while maintaining interpretability. The framework's ability to automate anomaly detection in large-scale datasets addresses scalability challenges faced by modern transportation TOCs.

By implementing the developed short ticketing detection algorithm, we can now precisely deploy revenue protection staff to stop passengers attempting to travel with short ticketing. This targeted approach also eliminates the need for random patrols, thereby optimizing the use of human resources. 
This research targets the location of stations with a high risk of short ticketing issues. However, the exact time period for each specific station is still to be discussed, which is not mentioned in the report.

\section{Introduction and Comprehensive Problem Analysis}

\subsection{Definition}

Short ticketing represents one of the common and persistent forms of fare evasion in modern railway systems, costing TOCs a tremendous amount of money annually in lost revenue. Unlike simple fare evasion where passengers travel without any ticket, short ticketing involves direct manipulation of the legitimate ticketing systems to achieve unauthorized travel at reduced cost.

\begin{figure}[htbp]
    \centering
    \includegraphics[width=0.8\linewidth]{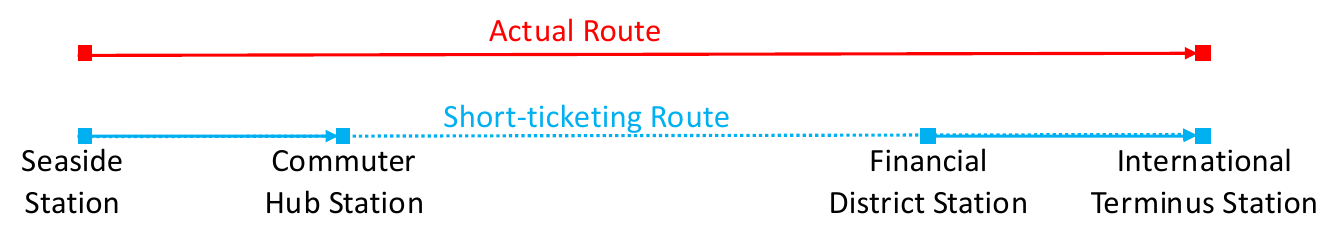}
    \caption{Example illustration of a short ticketing issue.}
    \label{fig:demo}
\end{figure}

This study begins with a practical illustration that demonstrates the complexity of this short ticketing mechanism. Consider a passenger traveling from Seaside Station to International Terminus Station via intermediate Commuter Hub Station and Financial District Station. Instead of purchasing a single through-ticket for the complete journey, the passenger purchases two separate tickets: Seaside Station to Commuter Hub Station, and Financial District Station to International Terminus Station The first ticket is used for entry at Seaside Station but not validated at the intermediate Commuter Hub Station. Similarly, the second ticket is used for entry at International Terminus Station but not validated at Financial District Station.
This scheme exploits several vulnerabilities in traditional railway ticketing systems:

\subsubsection{System Architecture Vulnerabilities}

\begin{itemize}
    \item Inconsistent validation requirements at intermediate stations
    \item Limited real-time integration between ticketing and gate systems
    \item Inadequate cross-referencing of entry/exit events with ticket validity
\end{itemize}

\subsubsection{Behavioural Exploitation Patterns}

\begin{itemize}
    \item Strategic selection of routes with minimal enforcement presence
    \item Timing travel to coincide with peak periods when manual checking is reduced
    \item Exploitation of complex multi-TOC route structures
    \item Manipulation of ticket validation requirements at interchange stations
\end{itemize}

\subsection{Problem Statement}

In the current stage, the short ticketing issue has only few methods to deal with. The most common solution could be seen as human detection, i.e., through ticket inspections by an inspector. However, such measures can only be observed on specific journey segments, and there remains a high probability of evading detection. This approach is also considered reactive, expensive, and struggles to provide a complete picture of the problem.

\section{Research Motivation and Objectives}
There are three fundamental objectives for this study. The primary detection goal is to develop robust methodologies to identify suspicious stations that may be systematically used for short ticketing activities.

The second goal is to establish quantitative frameworks for identifying high-risk stations to quantitatively optimize allocation of enforcement resources. Railway TOCs cannot monitor all stations continuously and simultaneously, making numerical operation allocation particularly important.

The third goal is automation and scalability requirements. Modern railway networks generate massive volumes of transaction data that are impossible for human experts to analyse, necessitating automated detection systems.

\subsection{Operational and Technical Challenges}
This research identified several challenges for developing an automatic self-supervised short ticket detecting algorithm.

\subsubsection{Data Quality and Integrity Issues}
Railway operational data suffers from numerous quality challenges that make short ticketing detection complicated. Incomplete records arise from unexpected scenarios such as system malfunctions. Missing values in critical fields such as entry/exit timestamps and station identifiers can confuse decision systems. 


\subsubsection{Station ID Inconsistencies}
Station ID inconsistencies result from factors such as multiple naming conventions and legacy system problems.

\subsubsection{Passenger Behavior Complexity}
Modern railway systems accommodate diverse passenger behaviours that make short ticketing difficult. Multiple entry/exit scenarios arise from complex route structures. Manual gate passes for accessibility, maintenance, or emergency situations further create false tracking of passenger behaviour.

Different ticket media types, from traditional physical tickets to contactless payments and mobile QR code tickets, create complex data structures and validation patterns. Each medium may produce different data artifacts that must be normalized and filtered for effective analysis.

\subsubsection{Absence of Labelled Training Data}

Unlike many machine learning applications, short ticketing detection in railway systems does not have well-defined labels for supervised learning approaches. Confirmed short ticketing cases only represent a very small fraction of actual short ticketing, and are further biased by current enforcement allocation. This limitation requires unsupervised detection methods that can identify suspicious patterns without relying on historical short ticketing labels.

\subsubsection{Feature Engineering Complexity}

Extracting meaningful behavioural patterns from raw operational data requires sophisticated feature engineering approaches. The dataset only contains the entry and exit status of a specific subset of the existing tickets. Thus, it requires well-designed feature extraction techniques.

\subsubsection{Computational Scalability Requirements}

Modern railway networks generate enormous volumes of transaction data requiring efficient processing algorithms. The framework must be able to handle tremendous daily transactions while providing near-real-time detection capabilities.

\section{Theoretical Foundation and Methodological Framework}

\subsection{The A/B/C/D Classification Paradigm}

We developed an innovative theoretical framework that transforms the only provided entry and exit data to information-rich features for short ticketing detection. This A/B/C/D paradigm represents each station interaction through four distinct roles.

A (Actual Entry): Physical entry events represent actual passenger presence at station entry points, typically recorded through automated gate systems. These events mean that the A stations are actually used as the start stations as the tickets describe.

B (Declared Destination): A station is classified as B station if it is the intended as the end of journey, but the ticket is never used at station B. This means there is no evidence that the passenger gets off at the destination the ticket indicates.

C (Declared Origin): A station is classified as C station if it is the intended begining of journey, but the ticket is never used at station C. This means there is no evidence that the passenger gets on at the start of the trip as the ticket indicates.

D (Actual Exit): Physical exit events represent actual passenger departure from stations.

This classification system enables identification of gaps between declared intentions (B and C) and actual behaviour (A and D), which is fundamental to detecting various forms of fare evasion and system abuse.

\subsection{Advanced Feature Engineering Architecture}
The framework employs sophisticated feature engineering to capture complex behavioural patterns that might indicates short ticketing:

\subsubsection{Fundamental Count Features}

Raw frequency counts provide baseline activity metrics for each station across all four roles. These counts establish the foundation for more sophisticated ratio and percentage calculations while providing intuitive measures of station activity levels. Count A, B, C, D stations respectively.

\subsubsection{Sophisticated Ratio Analysis}

Ratio features compare different aspects of station activity to identify imbalances characteristic of short ticketing behaviour:

\textbf{A-to-D Ratio:} Compares actual entries to actual exits, revealing stations where passenger inflow and outflow patterns deviate from expected normal pattern. Normal stations typically exhibit A-to-D ratios near 1.0.

\textbf{B-to-C Ratio:} Compares declared destinations to declared origins, identifying stations that serve disproportionately as origins or destinations in passenger declarations. Extreme imbalances may indicate systematic exploitation in short ticketing.

\textbf{Entry-Exit Difference Ratio:} Expresses the absolute difference between entries and exits as a fraction of total activity, providing a normalized measure of directional imbalance that accounts for varying station sizes.

\textbf{Critical Overlap Analysis:}
BC Overlap represents the most sophisticated and predictive feature in the framework, measuring the extent to which stations serve dual roles as both declared destinations and declared origins. This metric is calculated as the intersection of B and C roles divided by their union, providing a percentage indicating overlap intensity.
High BC overlap values (approaching 1.0) suggest that passengers frequently declare a station as both a destination and subsequent origin, which is characteristic of split-ticketing schemes where passengers purchase separate tickets for different journey segments.
BC Significance scales the overlap measure by total station activity, ensuring that high-volume stations with substantial overlap receive appropriate prioritization over low-volume stations with similar overlap percentages but minimal overall impact.

\subsection{Entropy and Distribution Analysis}
Shannon entropy calculations quantify the diversity of station usage patterns across the four roles.

Percentage distribution features normalize activity across roles, enabling comparison between stations of different sizes and activity levels. These features reveal whether stations primarily serve single functions (e.g., terminus stations) or multiple functions (potential fraud hubs).

\section{Unsupervised Learning Architecture and Implementation
Comprehensive Multi-Method Detection Strategy}
The framework employs four complementary unsupervised learning algorithms, each designed to capture different types of anomalous behavior patterns. This multi-expert approach ensures comprehensive coverage of potential short ticketing mechanisms while reducing the likelihood of false positives through consensus-based decision making.

\subsection{Isolation Forest Algorithm Analysis}
Isolation Forest operates on the principle that anomalous data points are easier to isolate from the main data distribution than normal points \cite{liu2008isolation}. The algorithm constructs ensemble decision trees through random feature selection and split point generation, measuring anomaly scores based on the path length required to isolate individual data points.
In the context of short ticketing detection, Isolation Forest proves particularly effective at identifying stations with unusual combinations of features that may not be apparent through individual feature analysis. For example, a station might exhibit normal entry/exit ratios and normal declared destination/origin ratios individually, but the combination of these features might be statistically improbable in legitimate scenarios.
The algorithm's strength lies in its ability to detect global outliers—stations that deviate significantly from overall network patterns. This capability is essential for identifying sophisticated fraud schemes that may appear normal when examined through limited feature sets but reveal anomalous patterns when considered.
\subsection{Local Outlier Factor (LOF) Implementation}
LOF addresses a fundamental limitation of global anomaly detection methods by focusing on local density deviations \cite{breunig2000lof}. The algorithm measures the local density of each data point relative to its neighbors, identifying points that exist in significantly sparser regions than their surrounding areas.
This approach is particularly valuable for short ticketing detection because legitimate station behavior varies significantly based on location, size, and operational characteristics. A behavior pattern that might be normal for a major station could be highly anomalous for a small suburban station. LOF's local analysis capability enables detection of context-appropriate anomalies.
The algorithm calculates local reachability density for each point and compares it to the densities of its k-nearest neighbors. Points with significantly lower densities than their neighbors receive high LOF scores, indicating local anomaly status.

\subsection{One-Class SVM Boundary Detection}
One-Class Support Vector Machine creates a decision boundary that encompasses the majority of normal data points in a high-dimensional feature space \cite{scholkopf1999support}. Points falling outside this boundary are classified as anomalies, making this approach particularly effective for detecting previously unknown fraud patterns.
The algorithm maps input features into a high-dimensional space using kernel functions, then constructs a hyperplane that separates normal data from the origin with maximum margin. This approach is especially valuable for short ticketing detection because it can identify novel patterns that may not resemble historically known anomalies.

\subsection{Mahalanobis Distance Statistical Analysis}
Mahalanobis distance provides a statistically grounded approach to anomaly detection by measuring how many standard deviations each data point lies from the multivariate mean while accounting for feature correlations and covariance structure \cite{mclachlan1999mahalanobis}.
This approach is particularly valuable for short ticketing detection because it considers the statistical properties of the entire dataset when evaluating individual stations. Unlike Euclidean distance, Mahalanobis distance accounts for the fact that different features may have different scales and may be correlated with each other.

\section{Advanced Weighting and Integration Framework}
\subsection{Evolution from Equal to Adaptive Weighting}
The framework initially employed equal 25\% weights across all four detection methods, treating each algorithm's contribution as equally valuable. However, analysis revealed that different methods sometimes identify similar anomaly types, leading to redundant detection and potential bias toward specific fraud patterns.
The evolution to adaptive weighting addresses this limitation through cross-correlation analysis between method rankings. Methods that produce highly correlated results receive reduced individual weights, while methods that provide unique perspectives receive enhanced weights.

\subsection{Cross-Correlation Analysis Implementation}
The correlation coefficient between method rankings is calculated using Spearman's rank correlation:
$\rho = 1 - (6\sum d^2) / (n(n^2 - 1))$
Where $d$ represents the difference between ranks assigned by two methods for each station, and n represents the total number of stations.
High correlation values $(\rho \ge 0.7)$ indicate that two methods largely agree on station rankings, suggesting redundancy. Low correlation values indicate that methods capture different aspects of anomalous behaviour, increasing their individual weights in the final score calculation.

\subsection{Dynamic Weight Assignment}
Adaptive weights are calculated through inverse correlation relationships:
\begin{equation}
    w_i = (1 - ac_i) / \sum(1 - ac_j)
\end{equation}
, where $ac$ means average correlation

This formulation ensures that methods with lower average correlations to other methods receive higher weights, promoting diversity in the detection ensemble.

\subsection{Rank-Based Score Integration}
The framework converts raw anomaly scores to ranks to address scale differences between methods. Each method ranks all stations from 1 (most anomalous) to N (least anomalous), enabling direct comparison and combination of results.
Final anomaly ranks are calculated as weighted averages:
\begin{equation}
    Final_Rank = \sum(w_i × Rank_i)
\end{equation}

These ranks are then normalized to produce intuitive anomaly scores ranging from 0 (completely normal) to 1 (most anomalous).

\begin{figure}
    \centering
    \includegraphics[width=0.6\linewidth]{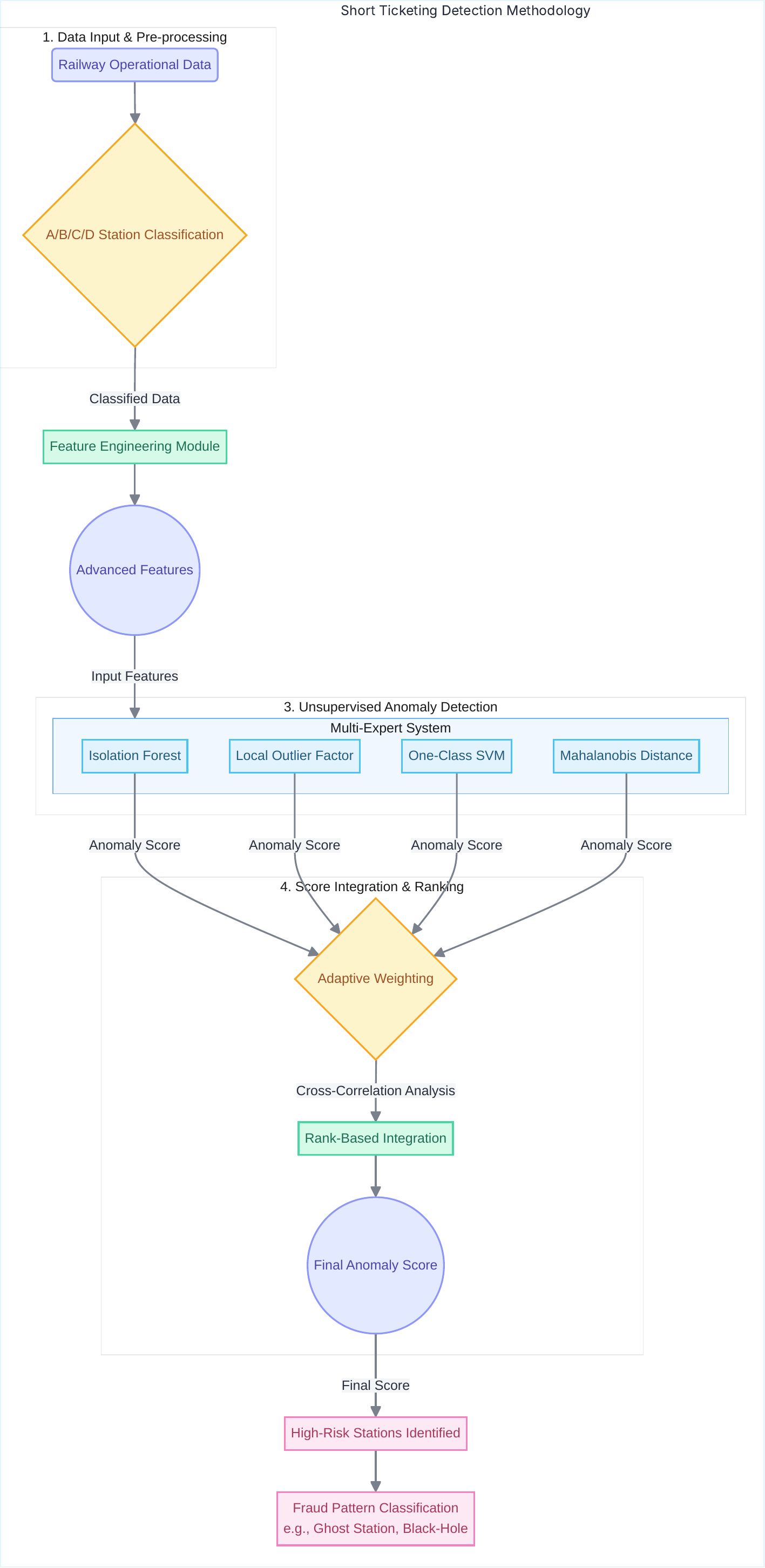}
    \caption{Figure on algorithm.}
    \label{fig:placeholder}
\end{figure}

\section{Comprehensive Short Ticketing Pattern Taxonomy}
\textbf{Ghost Station Behavior}

Predominance of entry-only or exit-only tickets; Near-zero legitimate through-journeys or transfers; Extreme; statistical deviation in multiple metrics; Low entropy indicating concentrated activity patterns.
 \\
\\
\textbf{Black-Hole Station Behavior}
Disproportionate exit-only ticket patterns; High entry/exit discrepancy ratios; Strong LOF detection indicating local anomaly status; Missing entry validation for high proportion of exits.
 \\
\\
\textbf{Fake-Origin Behavior}

High proportion of entry-only tickets; Strong One-Class SVM detection; Geographic positioning enabling route circumvention; Entry events without corresponding exit validation.
 \\
\\
\textbf{Function-Loss Behavior}

Extreme B/C ratio imbalances; Abnormal entry/exit loops;

\section{Comprehensive Results Analysis and Fraud Pattern Classification}

\subsection{Detailed Station-by-Station Risk Assessment}

\subsubsection{Tier 1 Critical Risk Stations (Anomaly Score $\ge$ 0.990)}

Airport Terminal Station (Anomaly Score: 1.000)
This station exhibits the most extreme anomalous behaviour identified in the entire network analysis. The perfect anomaly score of 1.000 indicates unanimous agreement across all detection methods that this station represents the highest short ticketing risk.
 \\
\\
\textbf{Behavioral Analysis}

\begin{itemize}
    \item Ghost Station Pattern: 63\% of tickets are entry-only, 37\% are exit-only, with zero legitimate through-journeys.
    
    \item Statistical Deviation: Mahalanobis distance of 364,255 represents unparalleled  deviation from normal patterns.

    \item Entry/Exit Imbalance: A-to-D ratio of 1.72 indicates 72\% more entries than exits

    \item Low Activity Diversity: Entropy score of 0.47 indicates highly concentrated activity patterns

    \item Zero BC Overlap: Complete absence of dual-role activity suggests systematic avoidance of normal transfer patterns
\end{itemize}

This pattern suggests systematic exploitation where passengers use the station as an entry point without legitimate exit validation, or exit point without proper entry validation. The extreme statistical deviation indicates this behaviour is unprecedented in the network and requires immediate investigation.

Downtown Station (Anomaly Score: 0.999)

The second-highest risk station demonstrates classic "Black-Hole" behavior patterns characteristic of systematic exit-point exploitation.
 \\
 \\
\textbf{Behavioral Analysis}

\begin{itemize}
    \item Exit Dominance: 62\% of tickets are exit-only, suggesting unauthorized departure procedures.
    
    \item Local Density Isolation: LOF score of 248,608 indicates extreme isolation from normal station behavior patterns.

    \item Entry/Exit Discrepancy: 24\% imbalance between entry and exit counts

    \item Missing Entry Validation: High proportion of exit events without corresponding entry records

\end{itemize}

This pattern suggests passengers are exiting through this station without proper entry validation, possibly through exploitation of gate malfunctions, manual override procedures, or coordination with staff. The extreme LOF score confirms this behaviour is anomalous even compared to other high-traffic terminal stations.

West Interchange Station (Anomaly Score: 0.996)

This station exhibits "Fake-Origin" patterns suggesting systematic exploitation as an artificial journey starting point.
 \\
 \\
\textbf{Behavioral Analysis}

\begin{itemize}
    \item Entry Dominance: 58\% entry-only tickets indicate disproportionate use as journey origin
    
    \item SVM Boundary Violation: Strong detection by One-Class SVM indicates behaviour outside normal boundaries

    \item Artificial Journey Starts: High proportion of entry events without corresponding exit validation

    \item Route Exploitation: Geographic location suggests potential use in routing longer-distance fare requirements
    
\end{itemize}

\section{Conclusions and Future Work}
In this study, we have developed a unsupervised multi-expert short ticketing system. We have identified several stations with severe short ticketing and analysis their abnormal pattern. 

Confidence in the results is high within the scope of the analyzed dataset, as the multi-expert framework successfully identified 30 high-risk stations showing highly unusual activity. We acknowledge, however, that the dataset covering 100 stations is a subset of the network, creating "boundary effects" where journeys starting outside this scope may be misidentified. The findings are also limited by potential data quality issues and complex passenger behaviours, such as manual gate passes, which can mimic fraudulent patterns.

Future work will focus on addressing these gaps to improve the model. Expanding the dataset to a larger portion of the network will reduce boundary effects. Furthermore, accuracy can be enhanced by integrating higher-fidelity data sources, such as barcode ticket information and gate status logs, to better distinguish true anomalies from operational artifacts. In addition, the high-risk time period of the short ticketing issue should be individually considered, which could provide a more accurate and precise solution.

\section{Acknowledgement}
We would like to express our sincere gratitude to Cubic Transportation Systems \cite{Cubic.com} for their valuable support throughout this research project. Their collaboration, industry expertise, and provision of the essential railway dataset were critical to the development and validation of the short ticketing detection framework presented in this report.

\cleardoublepage
\printbibliography

\end{document}